\documentclass[iop]{emulateapj}
\usepackage[english]{babel}
\usepackage{times}
\def\mincir{\raise -2.truept\hbox{\rlap{\hbox{$\sim$}}\raise5.truept \hbox{$<$}\ }}
\def\mincireq{\hbox{\raise0.5ex\hbox{$<\lower1.06ex\hbox{$\kern-1.07em{\sim}$}$}}}
\def\magcir{\raise-2.truept\hbox{\rlap{\hbox{$\sim$}}\raise5.truept \hbox{$>$}\ }}
\def\gr{\kern 2pt\hbox{}^\circ{\kern -2pt K}} 

\def\_{\thinspace}

\shorttitle{SED vs. activity in blazars }
\shortauthors{Mankuzhiyil et al.}

\begin{document}

\title{EMITTING ELECTRONS AND SOURCE ACTIVITY\\ IN MARKARIAN\,501}

\author{Nijil\,Mankuzhiyil\altaffilmark{1,2}}
\affil{INFN, Trieste, Italy}

\author{Stefano\,Ansoldi\altaffilmark{3,4}}
\affil{International Center for Relativistic Astrophysics (ICRA), Rome, Italy}

\author{Massimo\,Persic\altaffilmark{3}}
\affil{INAF-Trieste, via G.\,B.\,Tiepolo 11, I-34143 Trieste (TS), Italy}

\author{Elizabeth\,Rivers, Richard\,Rothschild }
\affil{Center for Astrophysics and Space Sciences, University of California at San Diego, \\
9500 Gilman Drive, La Jolla, CA 92093-0424, USA}

\author{Fabrizio\,Tavecchio}
\affil{INAF-Brera, via E.\,Bianchi 46, I-23807 Merate (LC), Italy}

\altaffiltext{1}{and Dipartimento di Chimica Fisica e Ambiente, Universit\`{a} di Udine, via delle Scienze 208, I-33100 Udine (UD), Italy}
\altaffiltext{2}{and Inter-University Consortium for Space Physics, Torino, Italy}
\altaffiltext{3}{and INFN, Trieste, Italy}
\altaffiltext{4}{and Dipartimento di Matematica e Informatica, Universit\`{a} di Udine, via delle Scienze 206, I-33100 Udine (UD), Italy}


\begin{abstract}
We study the variation of the broad-band spectral energy distribution (SED) of the
BL\,Lac object Mrk\,501 as a function of source activity, from quiescent to flaring.
Through $\chi^2$-minimization we model eight simultaneous SED datasets with a
one-zone Synchrotron-Self-Compton (SSC) model, and examine how model parameters vary
with source activity. The emerging variability pattern of Mrk\,501 is complex, with
the Compton component arising from $\gamma$-e scatterings that sometimes are (mostly)
Thomson and sometimes (mostly) extreme Klein-Nishina. This can be seen from the variation of
the Compton to synchrotron peak distance according to source state. The underlying
electron spectra are faint/soft in quiescent states and bright/hard in flaring states.
A comparison with Mrk\,421 suggests that the typical values of the SSC parameters are
different in the two sources: however, in both jets the energy density is particle
dominated in all states.
\end{abstract}


\section{Introduction}

The fueling of supermassive black holes, hosted in the cores of most galaxies, by infalling matter
is thought to produce the spectacular activity observed in AGNs. In $\mincir$10\% of cases powerful
collimated jets shoot out in opposite directions at relativistic speeds. If a relativistic jet is
viewed at a small angle to its axis, the observed emission is amplified by relativistic beaming
(Doppler boosting and aberration; see Urry \& Padovani 1995). Sources, whose boosted jet emission
dominates the observed emission (blazars
\footnote{
    Extreme blazars, whose thermal emission is intrinsically weak
    (i.e., no emission lines in their spectra), are called BL\,Lac
    objects.
         }
), are crucial to studying the physics of relativistic extragalactic jets.

The jets' origin and nature are unclear. They may be (kinetic or electromagnetic) flows that dissipate some
of their energy in moving regions associated with internal or external shocks. This complex picture is
approximated, for the purpose of modelling the observed emission, with one (or more) relativistically-moving,
magnetized, homogeneous plasma region (blob), where a time-varying non-thermal population of electrons emit
radiation (e.g., Maraschi et al. 1992). The latter is synchrotron radiation and its comptonized (IC) counterpart,
that peak at (respectively) IR/X-ray and GeV/TeV frequencies {[in low/high-frequency-peaked BL\,Lac sources
(LBL/HBL)]} and show correlated luminosity and spectral changes. In the case of HBLs one same population of
electrons in the blob generates the low-energy (IR to X-ray) synchrotron electrons and Compton-(up)scatters
them to high energies [Synchrotron-Self-Compton (SSC) mechanism], with no external sources of soft photons.

The recent availability of simultaneous broad-band SEDs (mostly for nearby BL\,Lac objects) has enabled addressing
the important issue of how the emission changes as a function of the source's global level of activity: i.e., given
an emission model that fits the data, it is should be examined what model parameters are correlated with source
activity. Another requirement concerns using a full-fledged $\chi^2$-minimization procedure instead of the "eyeball"
fits still most commonly used in the literature, in order to obtain unbiased results.

In a previous study of this type, Manzkuzhiyil et al. (2011) examined nine SEDs of the {HBL}
object Mrk\,421 ($z$=0.031). In this follow-up paper we analyze eight SEDs of the HBL source
Mrk\,501 ($z$=0.034), applying the same one-zone SSC emission model and $\chi^2$-minimization procedure (Sect.2
and 3.1, respectively) to the datasets described in Sect.3.2. The results are presented and discussed in Sect.4,
and summarized in Sect.5.

\section{BL\,Lac SSC emission}

Following Mankuzhiyil et al. (2011), to describe the HBL broad-band emission we use a one-blob SSC model
(Tavecchio et al. 1998, later T98; Maraschi \& Tavecchio 2003). This adequately describes broad-band SEDs of most HBLs
(Tavecchio et al. 2010) and, for a given source, both its ground and excited states (Tavecchio et al. 2001;
Tagliaferri et al. 2008). Convincing evidence for the SSC model is the X-ray/VHE-$\gamma$-ray variability
correlation (e.g., Fossati et al. 2008): since in the SSC model the emission in the two bands is produced
by the same electrons, a strict X-ray/$\gamma$-ray correlation is expected
\footnote{The rarely occurring ``orphan'' TeV flares, that are not accompanied by
          variations in the X-ray band, may arise from small, low-$B$, high-density
      plasma blobs (Krawczynski et al. 2004).} .

In this work, for simplicity we use a one-zone SSC model, assuming that the entire SED is produced
within a single homogeneous region of the jet. Although generally adequate to reproduce HBL SEDs,
one-zone models have troubles explaining some specific features of blazar TeV emission. In particular,
while very large Doppler factors are often required in a one-zone model, radio VLBI observations hardly
detect superluminal motion at parsec scale (e.g., Piner et al. 2010; Giroletti et al. 2006). This led
to the proposal of the existence of a structured, inhomogeneous and decelerating emitting jet
(Georganopoulos \& Kazanas 2003; Ghisellini et al. 2005). Inhomogeneous (two-zone) models (e.g., Ghisellini
\& Tavecchio 2008) have been also invoked to explain the ultra-rapid variability occasionally observed in TeV
blazars (e.g., Aharonian et al. 2007; Albert et al. 2007).

In the one-zone SSC model the emitting plasma is contained in a spherical blob of radius $R$ in
relativistic motion (described by a bulk Lorentz factor $\Gamma$) along the jet at an angle $\theta$
w.r.t. the line of sight to the observer, so that special relativistic effects are cumulatively
described by the relativistic Doppler factor, $\delta=[\Gamma(1-\beta\,{\rm cos}\,\theta)]^{-1}$.
The blob is filled with a homogeneous tangled magnetic field with intensity $B$ and by a population
of relativistic electrons of density $n_e$, whose spectrum is described by a broken power-law (PL)
function of the electron Lorentz factor $\gamma$,
\begin{eqnarray}
{N_e(\gamma)} ~=~ K_e~ \times ~
\left\{
\begin{array}{ll}
\gamma^{-n_1}                           & \mbox{~~~~~$\gamma_{\rm min} \leq \gamma \leq \gamma_{\rm br}$} \\
\gamma_{\rm br}^{n_2-n_1} \gamma^{-n_2} & \mbox{~~~~~$\gamma_{\rm br} < \gamma \leq \gamma_{\rm max}$} \,.
\end{array}
\right.
\label{eq:el_spectr}
\end{eqnarray}
(Of course, $n_e = \int_{\gamma_{\rm min}}^{\gamma_{\rm max}} N_e(\gamma)\, {\rm d}\gamma$\,.) This
purely phenomenological choice is motivated by the observed shape of the humps in the SEDs (e.g., T98).

In shaping the VHE\,$\gamma$-ray part of the spectrum it is important: {\it (i)} to use the {\it full}
Klein-Nishina (K-N) cross section (see Maraschi \& Tavecchio 2003) in the T98 model; and {\it (ii)} to
correct $\gtrsim 50\,$GeV data for Extragalactic Background Light (EBL) absorption -- as a function of
photon energy and source distance (e.g., Mankuzhiyil et al. 2010): here we use the popular Franceschini
et al. (2008) EBL model.

It is {useful} to sample the SED around {\it both} peaks, because the synchrotron emissivity contains
degeneracy between $B$ and $K_e$ (i.e., $j_{\rm s} \propto K_eB^{(n+1)/2}$, with $n=2\,\alpha+1$ where a
$n$ is the electron's spectral index and $\alpha$ is the measured synchrotron emission profile) that cannot
be lifted without knowledge of the IC peak, which provides the desired second equation (e.g., $j_{\rm IC}
\propto K_e^2 B^{(n+1)/2}$). Therefore, in general only knowledge of observational quantities related to both
SED humps enables {reliable} determination of all SSC parameters. In particular, the one-zone SSC model
can only be fully constrained by using simultaneous broad-band observations. Of the 9 free parameters of the
SSC model, 6 specify the electron energy distribution ($n_{\rm e}$, $\gamma_1$, $\gamma_{\rm br}$, $\gamma_2$,
$n_1$, $n_2$), and 3 describe the global properties of the emitting region ($B$, $R$, $\delta$). Some
observational quantities are linked to the SSC model parameters: the slopes, $\alpha_{1,2}$, of the
synchrotron bump on either side of the peak are linked to $n_{1,2}$ (whence the phenomenological choice
of a broken-PL electron spectrum); the synchrotron and IC peak frequencies, $\nu_{\rm s,IC}$, and
luminosities, $L_{\rm s,IC}$, are linked with $B$, $n_{\rm e}$, $\delta$, $\gamma_{\rm br}$; finally,
the minimum variability timescale $t_{\rm var}$ provides an estimate of the source size through $R \,
\mincir \, c t_{\rm var} \delta /(1+z)$.

\begin{figure*}
\begin{center}
\includegraphics[width=17.2cm]{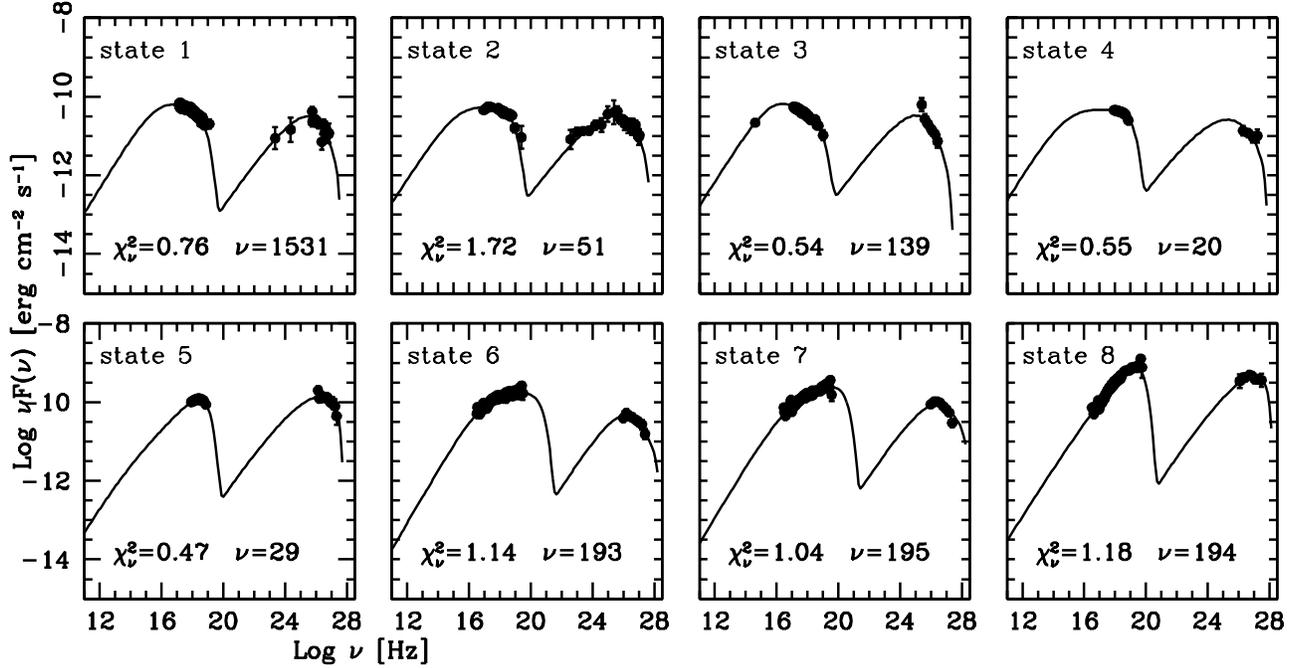}
\caption{%
Eight SEDs of Mrk\,501. See Table\,1 for details.
}
\end{center}
\label{fig:SED_fit}
\end{figure*}

\section{$\chi^2$-minimization}

\subsection{The code}

We start by setting $\gamma_{\rm 1} = 1$, as usually assumed in the literature. A search for the best
values (and associated uncertainties) of the remaining 8 free parameters is performed by best-fitting
our {chosen} SSC model to the SED data: the corresponding $\chi^2$-minimization is based on the
Levenberg--Marquardt method -- an efficient standard for non-linear least-squares minimization that
{dynamically}
interpolates between two different minimization approaches, the inverse Hessian method and the steepest
descent method.

A description of how the code works is given in Man\-ku\-zhi\-yil et al. (2011) and Ansoldi (2012). Let's stress two
important points in our implementation. {\it (i)} The T98 model only returns {the} numerical sample of the
SSC SED. On the other hand, at each step of the loop the calculation of $\chi^{2}$ requires evaluating
the SED at all the {\it observed} frequencies. If the model function is known analytically, these
evaluations are a straightforward algebraic process. In our case, instead, we know the model function
only through a numerical sample coming from the implementation of the T98 model, and in general these
sampled points do not {include} the observed points. The latter, however, typically fall between
two sampled points, which -- given the dense sampling of the model SED -- allows us to use interpolation
to approximate the {SED value at the} observed point with negligible uncertainties. {\it (ii)} The Levenberg--Marquardt
method requires the calculation of the partial derivatives of $\chi^{2}$ with respect to the 8 floating
SSC parameters. Contrary to the case when knowledge of the analytical form of the model function enables
all derivatives to be obtained {in closed form}, in our case the derivatives {are} numerically obtained by
evaluating the increment of $\chi^2$ with respect to a sufficiently small, dynamically adjusted increment
of each parameter. As this procedure could have led to some computational inefficiency due to the recurrent
need to evaluate the SED {from the model} at many finely-spaced points in parameter space
(a CPU-time consuming operation),
we set up a specific algorithm to minimize the number of calls to T98 across different iterations.

Because of the subtleties in non-linear fits we optimized the procedure to identify the parameters
uncertainties for reliability and statistical significance. The standard approach (uncertainties as the
square-root of the covariance matrix diagonal elements) depends on a reliable quadratic approximation to
the $\chi^2$-surface around the minimum: otherwise uncertainties can be underestimated/overestimated. To
avoid this we directly solve for them using (we call the SSC parameters $p_{i}$ here)
\[
    \chi ^{2} ( p _{1} ^{(\mathrm{min})} , \dots , p _{i} , \dots , p _{8} ^{(\mathrm{min})})
    =
    \chi ^{2} _{\mathrm{min}} + \alpha,
    \quad
    i = 1, 2, \dots, 8,
\]
where $\alpha = 1, 2.71, 6.61$ realize a $68\%$, $90\%$ and $99\%$ confidence level, respectively
\footnote{Our choice is the standard $90\%$ confidence level.}
and $\chi ^{2} _{\mathrm{min}} = \chi ^{2} ( p _{1} ^{(\mathrm{min})} , \dots , p _{8} ^{(\mathrm{min})})$
is the chi-square at the optimal estimation for the free parameters. Starting from the minimum point, the
above equations are solved twice for each parameter $p _{i}$, keeping the other seven fixed to their optimal values.
We first add to the floating parameter positive increments, until we reach a value $p _{i} ^{(\mathrm{right})}
> p _{i} ^{(\mathrm{min})}$ at which $\chi ^{2} ( p _{1} ^{(\mathrm{min})} , \dots ,p _{i} ^{(\mathrm{right})},
\dots , p _{8} ^{(\mathrm{min})}) > \chi ^{2} _{\mathrm{min}} + 2.71$. Then, $p _{i} ^{(\mathrm{min})}$ and
$p _{i} ^{(\mathrm{right})}$ represent what is called a \emph{bracketing} of the solution $p _{i} ^{(\alpha_
{\mathrm{r}})}$ at which $\chi^{2} ( p_{1} ^{(\mathrm{min})} , \dots , p_{i} ^{(\alpha_{\mathrm{r}})} , \dots ,
p _{8} ^{(\mathrm{min})}) = \chi ^{2} _{\mathrm{min}} + 2.71$. Given the bracketing, we numerically solve for
$p _{i} ^{(\alpha _{r})}$ using Ridders' method
\footnote{Ridders' method (Ridders 1979) only requires evaluation of the function which
      is searched for a zero, not of the function derivatives: it is thus well suited
      to our case, in which calculation of $\chi ^{2}$ (and thus model function)
      derivatives is computationally demanding. The order of Ridders' method is $\sqrt{2}$
      (Press 1992), so it is superlinear and competitive with more refined approaches.}
(Press 1992) and define the right uncertainty associated to $p _{i}$ as $\sigma_{i} ^{(\mathrm{right})} =
p_{i} ^{(\alpha_{r})} - p_{i}^{(\mathrm{min})}$. We then proceed similarly to the left of $p_{i} ^{(\mathrm{min})}$,
we identify $p_{i}^{(\alpha _{l})}$ so that $\chi^{2} ( p_{1}^{(\mathrm{min})} , \dots , p_{i} ^{(\alpha_{l})}
, \dots , p_{8}^{(\mathrm{min})}) = \chi^{2}_{\mathrm{min}} + 2.71$ and the left uncertainty is then $\sigma_{i}^
{(\mathrm{left})} = p _{i} ^{(\mathrm{min})} - p _{i} ^{(\alpha _{l})}$. The final results for the parameters are
then estimated as $p_{i} ^{(\mathrm{min})} {}^{+\sigma_{i} ^{(\mathrm{right})}}_{-\sigma _{i} ^{(\mathrm{left})}}$
at the $90\%$ confidence level.

\subsection{The data sets}

Searching the literature for simultaneous broad-band SED data sets of Mrk\,501 returned the following:
\smallskip

\noindent
$\bullet$
{\it State\,1} data were collected by {\it Suzaku} (March 23-25, 2009), by MAGIC and VERITAS (March 23 and March 24 2009,
respectively), and by {\it Fermi}/LAT (all throughout the campaign) -- see Acciari et al. (2010).
\smallskip

\noindent
$\bullet$
{\it State\,2} data were taken from the most intensive multi-wavelength campaign ever conducted on Mrk\,501 (Abdo et al.
2011). Swift, {\it R}XTE, {\it Fermi}/LAT, MAGIC and VERITAS data are used in this work. The X-ray data were obtained
during Mar\,15-Aug\,1, 2009. Fermi observed the source in the same time window, but data taken in May were omitted in this
work due to a variable high-energy flux. MAGIC observations were carried out between Apr\,27-May\,9, 2009: however, due to bad
weather and a shutdown for hardware upgrade, the observation days were significantly reduced from the original schedule.
VERITAS observations took place between Mar\,17-Jun\,22, 2009: but data from May\,1-4 were dropped due to high variability.
\smallskip

\noindent
$\bullet$
{\it State\,3} data come from a multi-frequency campaign in July 2006 (Anderhub et al. 2009). KVA (optical) and MAGIC
(VHE\,$\gamma$-ray) observations were carried out on July 18, 19, and 20 -- plus a few tens of minutes in August and
September 2006. Suzaku observations were carried out on July 18 and 19, 2006. During the whole campaign, no significant
flux variability was observed in any of the three frequency ranges.
\smallskip

\noindent
$\bullet$
{\it State\,4} and {\it State\,5} data come from a campaign carried out by {\it R}XTE and HEGRA in June 1998: State\,4 uses
observations from June\,15-26, 1998, whereas State\,{5} uses data taken during a higher-activity state on June\,27-28,
1998 (Sambruna et al. 2000).
\smallskip

\noindent
$\bullet$
{\it States 6, 7, 8} are based on a multi-frequency campaign, involving CAT and {\it Beppo}Sax, from Mar-Oct 1997. The
X-ray data of states 6 and 7 were obtained on April 7 and 11, respectively. The VHE\,$\gamma$-ray fluxes used in State\,6
and 7 represent average emissions of the low and intermediate states observed during the campaign
($\mathrm{flux} < 1.2 \times 10^{-10}$\,cm$^{-2}$\,s$^{-1}$ and
$1.2 \times 10^{-10}$\,cm$^{-2}$\,s$^{-1} < \mathrm{flux} < 5.3 \times 10^{-10}$\,cm$^{-2}$\,s$^{-1}$, respectively).
State\,8 comprises X-ray and VHE\,$\gamma$-ray data obtained during the giant flare of April 16, 1997 (Djannati-Ata\"{\i}
et al. 1999).

\section{Results}

Best-fit SSC models of simultaneous blazar SEDs are crucial to measuring the SSC parameters that describe
the emitting region. The emerging best model is {obtained by thoroughly
searching} for the absolute $\chi^2$ minimum. Our procedure ensures there is no obvious bias affecting
the resulting best-fit SSC model.

\begin{figure*}
\begin{center}
\includegraphics[width=17.2cm]{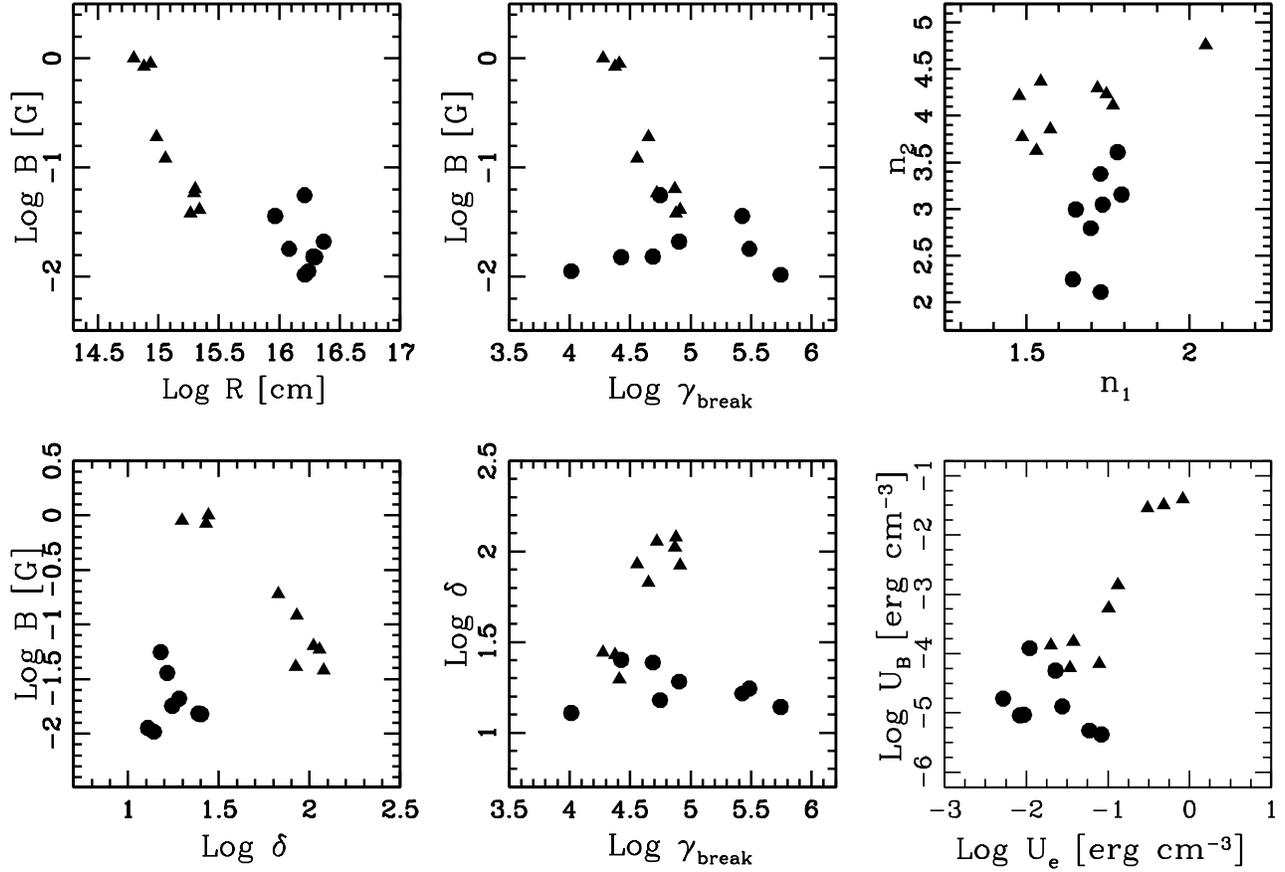}
\caption{%
Variations of the plasma blob's radius ($R$), magnetic field
($B$), and Doppler factor ($\delta$), and of the electrons'
spectral indices (n$_1$, n$_2$) and spectral break ($\gamma_
{\rm br}$). Also shown are the variations of the magnetic and
electron energy density ($U_B$ and $U_e$, respectively).
Circles and triangles denote Mrk\,501 and Mrk\,421, respectively.
{Uncertainties for Mrk\,421 (Mankuzhiyil et al. 2011) have been estimated again
with the method described in this paper to make them consistent with those of Mrk\,501.}
}
\end{center}
\label{fig:SSC_param}
\end{figure*}

The SED data and their best-fit SSC fit parameters are reported in Table\,1. The corresponding fits are
plotted in Fig.~\ref{fig:SED_fit}. The 90\% error bars reported in Table\,1 result from a detailed study
of the $\chi^2$ profiles around the minima corresponding to each parameter (see Sect.3.2), and in general
they turn out asymmetric. Within the available statistics (8 data points), source activity (measured
as the total luminosity of the best-fit SSC model
\footnote{
    The (isotropic) bolometric luminosities used in these plots have been obtained directly
    from the model SED. After determining the numerical approximation to the
    SED $\log [\nu F(\nu)]$, with the parameters being fixed at their best values
    obtained with the previously described minimization procedure, we have performed
    $L = 4 \pi D_L^2 \int_{\nu _{\mathrm{min}}} ^{\nu_{\mathrm{max}}} F (\nu) d \nu$, with
    $D_L$ the luminosity distance, and
    $\nu_{\mathrm{min}}$, $\nu_{\mathrm{max}}$ set at 2.5 and 0.75 decades, respectively,
    below the synchrotron peak and above the Compton peak. In this way we make sure
    to perform the integral over all the relevant frequencies in a way that is
    independent from {the} location of these peaks. (This procedure was also used in
    Mankuzhiyil et al. 2011 to calculate the luminosity of Mrk\,421, in spite of a typo
    that makes the description inaccurate.)}
seems to be most correlated with the electron spectrum normalization, $K _{\mathrm{e}}$ (see Fig.~\ref{fig:el_spectrum}),
and with the high-energy power-slope, $n_2$: both quantities vary with luminosity, in the sense that an
increase of luminosity means an increase of the electron density and its high-energy fraction.

In more detail, from Table\,1 -- within the low statistics involved -- we can see that:
\smallskip

\noindent
{\it (i)} in quiescence (states 1, 2, 3) apparently uncorrelated variations of the SSC parameters produce
SEDs of constant (isotropic) luminosity;
\smallskip

\noindent
{\it (ii)} in the sequence of June 1998 (states 4, 5) a $\sim$3-fold increase of luminosity is accompanied by
a 3-fold increase of $K_e$ and by a substantial electron spectral hardening -- as well as by a 2-fold
decrease of $\delta$;
\smallskip

\noindent
{\it (iii)} in the Giant Flare of 1997 (states 6, 7, 8) the 4-fold increase of luminosity is accompanied by
increments of $K_e$ and $\gamma_{\rm br}$ by factors of, respectively, 5 and 2, and by a strong hardening
of the electron spectrum -- as well as by a decrease of $\delta$ that eventually amounts to a $\sim$2-fold
decrease in relativistic boosting. Along this sequence the emission radius $R$ increases, and the magnetic
field $B$ decreases, such that the quantity $BR^2$ -- i.e., the magnetic flux across the boundary of the
emitting region -- appears to be approximately conserved;
\smallskip

\noindent
{\it (iv)} although in most emission states considered in this paper $n_2 - n_1 \gg 0$, in states 8 and 5,
i.e. in the flaring culminations of the April 1997 and June 1998 periods of activity, $n_2$ has decreased
such that $n_2 - n_1 \sim 0.5$: i.e., the electron spectrum, generally assumed to be double-PL, tends to
look like a single-PL spectrum during flares.

In general, we find that the kinetic energy density of relativistic electrons, $U_e = m_e c^2 \int_
{\gamma_{\rm min}}^{\gamma_{\rm max}} \gamma\, N_e(\gamma)\, {\rm d}\gamma$ which is a combination
of $K_e$, $n_1$, $n_2$, $\gamma_{\rm min}$, $\gamma_{\rm br}$, and $\gamma_{\rm max}$, is strongly correlated
with the blob rest-frame luminosity (Fig.~\ref{fig:el_spectrum}).

\subsection{Electron injection spectrum}

The large variation in luminosity and SED shape displayed by Mrk\,501 between flaring and non-flaring states
is matched by a concomitant change in the electron spectrum, $N_{\rm e}(\gamma)$ (see Fig.~\ref{fig:el_spectrum}):
\smallskip

\begin{figure*}
\begin{center}
\includegraphics[width=17.2cm]{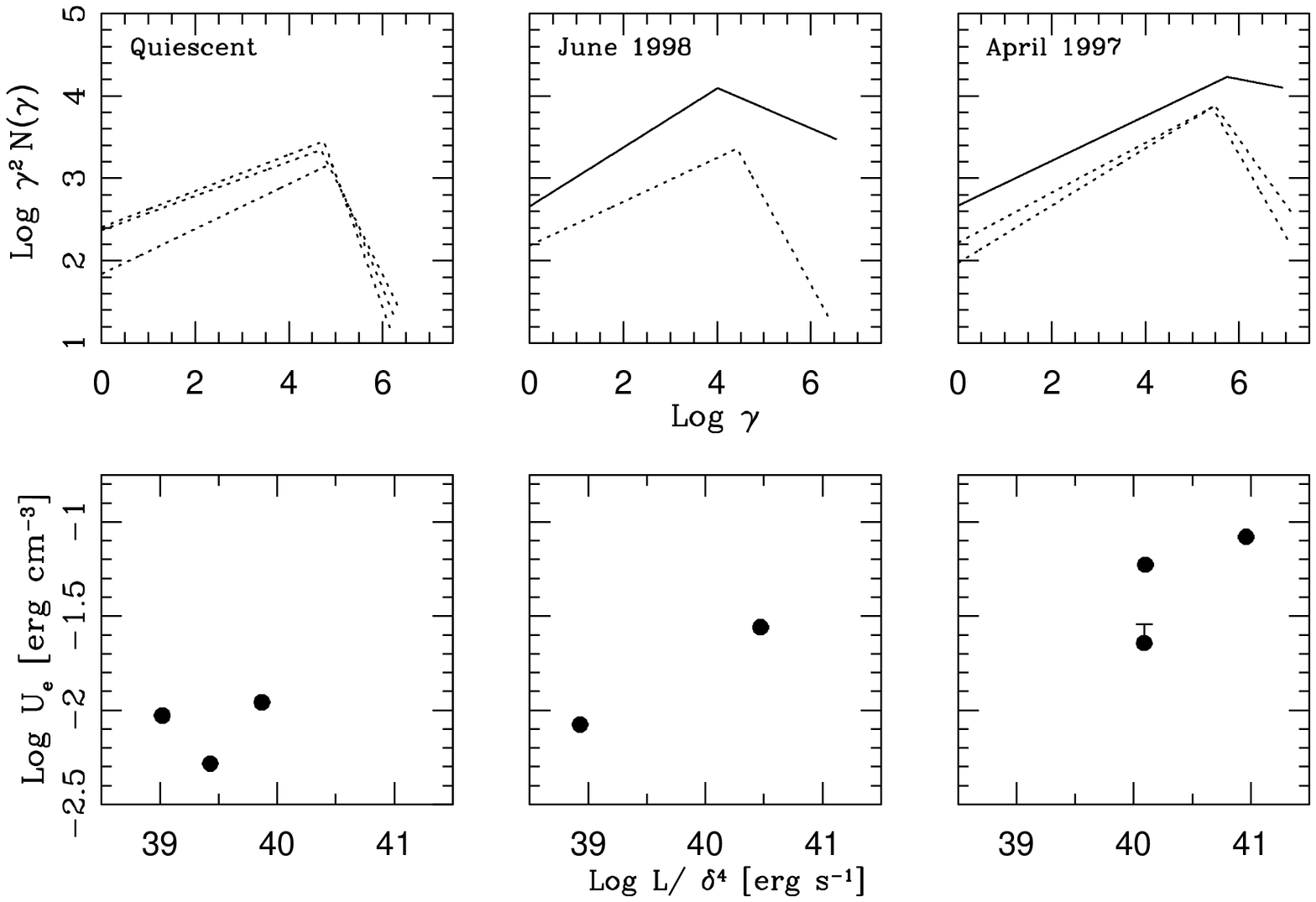}
\caption{
Electron spectra ({\it top}) and energy densities ({\it bottom}) in Mrk\,501's jet during:
({\it left}) quiescent times (uncorrelated states 1, 2, 3);
({\it middle}) June 1998 event (time-ordered states 4, 5);
({\it right}) April 1997 event (time-ordered states 6, 7, 8).
{The electron spectra, $N(\gamma)$, are plotted here multiplied by $\gamma^2$
in order to better illustrate elecrons of which energy carry the bulk of the kinetic energy.}
In the assumed one-blob SSC model, the quantity $L \delta^{-4}$ represents the bolometric SSC model
luminosity corrected for the effects of relativistic beaming (see Urry \& Padovani 1995): i.e.,
$L \delta^{-4}$ represents the blob-frame isotropic SSC luminosity.  }
\label{fig:el_spectrum}
\end{center}
\end{figure*}

\noindent
{\it (i)} In the three quiescent states (Fig.~\ref{fig:el_spectrum}, left) the spectra are clearly double-PL
($\Delta n \simeq 1.6$) with similar $\gamma_{\rm br} \simeq 6 \times 10^4$ and normalization $K_e$ ranging from
$\sim$70\,cm$^{-3}$ to 230{\,cm$^{-3}$} to 255{\,cm$^{-3}$}.
\smallskip

\noindent
{\it (ii)} Along the sequence of the two states of 1998 June 15-26 and 27-28, the spectrum hardens ($\Delta n$
from 1.32 to 0.61) and brightens ($K_e$ increases by a factor of 3) (Fig.~\ref{fig:el_spectrum}, center).

\noindent
{\it (iii)} Along the sequence of the three states (6,7,8) from 1997 April 7, 11, 16 the spectra become harder
in shape and higher in normalization as the luminosity rises (Fig.~\ref{fig:el_spectrum}, right). On April 7 the
normalization is lowest ($K_e \simeq 90$ cm$^{-3}$ and the spectral break is highest ($\Delta n \equiv n_2 - n_1 =
1.35$). By April 11 the normalization is intermediate ($K_e \simeq 165$ cm$^{-3}$) and the spectral break has
decreased ($\Delta n = 1.10$). On April 16, at the peak of the flare, the normalization is highest ($K_e \simeq
465$ cm$^{-3}$) and the spectral break has almost disappeared ($\Delta n = 0.38$).

The implied temporal pattern is quite clear. At the onset of the flare, the electron spectrum has a distinctly
double-PL slope and relatively low normalization. As the flare goes on, more and more newly accelerated electrons
are injected, and the resulting spectrum is a combination of the ``old'', double-PL spectrum and the ``new'', single-PL
injection spectrum. At the peak of the flare the injection spectrum dominates the emission. Correspondingly, the
electron energy density ($U_e$) shows an increasing trend with luminosity and, within a correlated sequence of
time-ordered states, with the development of the flare (see Fig.~\ref{fig:el_spectrum}).

\subsection{Thomson versus extreme K-N scattering regimes}

We can check whether the elementary scattering processes that give rise to the IC component of the SED
occur mainly in the Thomson or {extreme} K-N regime by studying the logarithmic distance between the synchrotron
and IC peaks. At its simplest, the argument goes as follows. In the electron's {rest} frame (primed quantities),
by scattering off the electron a photon with initial energy $\epsilon^\prime$ goes off at a scattering
angle $\theta_1^\prime$ and an energy $\epsilon_1^\prime$ given by
\begin{eqnarray}
\epsilon_1^\prime = {\epsilon^\prime \over 1 + (\epsilon^\prime /m_ec^2) \, (1-{\rm cos}\theta_1^\prime) }
\label{eq:scatter1}
\end{eqnarray}
(e.g., Blumenthal \& Gould 1970). In the lab system the energy is
\begin{eqnarray}
\epsilon_1 \,\, = \,\, \gamma \epsilon_1^\prime \,[1+\beta\, {\rm cos}(\pi-\theta_1^\prime)] \,\, = \,\,
\gamma \epsilon_1^\prime \,(1- \beta {\rm cos}\theta_1^\prime) \,,
\label{eq:scatter2}
\end{eqnarray}
so $\epsilon_1 \sim \gamma \epsilon_1^\prime$. Therefore, from Eq.(\ref{eq:scatter1}) we see the
two limiting cases:
\smallskip
\begin{figure}
\begin{center}
\includegraphics[width=8.4cm]{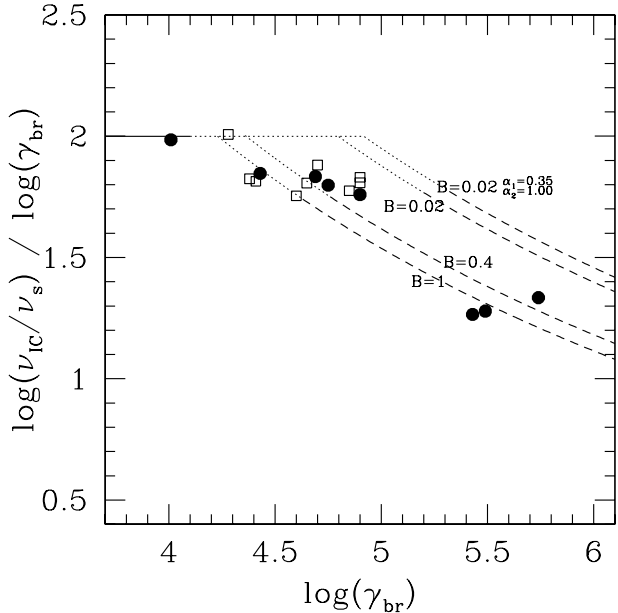}
\caption{Logarithmic distance between the IC and synchrotron peaks as a function of the electrons'
spectral break energy for the 8 SEDs of Mrk\,501 (filled circles) considered in this paper and the
9 SEDs of Mrk\,421 studied in Mankuzhiyil et al. (2011; empty squares). The thick solid and dashed
lines correspond to the Thomson ($h \nu /m_ec^2 \rightarrow 0$) and extreme K-N ($h \nu /m_ec^2 \rightarrow \infty$)
limits, respectively. In calculating the extreme K-N curves magnetic field values
of $B=0.02$\,G (appropriate for Mrk\,501, see Table\,1), and $B=0.4\,$G and 1\,G (appropriate for
Mrk\,421, see Mankuzhiyil et al. 2011) were assumed. (In all cases, when not indicated otherwise, we
used $\alpha_1=0.5$, $\alpha_2=1.5$.) Thin lines represent the extrapolation of the two limiting
regimes into the $h \nu \sim m_ec^2$ region. In the extreme K-N limit, the two peak frequencies are
closer than in the Thomson limit.
}
\end{center}
\label{fig:Thomson_KN}
\end{figure}

\noindent
{\it (i)} in the low-frequency limit $\epsilon^\prime \ll m_ec^2$ (Thomson regime), it is $\epsilon_1^\prime \sim
\epsilon^\prime$. The energy transformation from the lab frame to the electron frame, $\epsilon^\prime =  \gamma
\epsilon \,(1- \beta {\rm cos}\theta)$, implies $\epsilon^\prime \simeq \gamma \epsilon$, so $\epsilon_1 \simeq
\gamma^2 \epsilon$. In the SSC framework $\epsilon_1 = h \nu_{\rm IC}$ and $\epsilon = h \nu_{\rm s}$, and the
low-frequency condition writes $\gamma \ll m_ec^2 / (h \nu_{\rm s})$; since $\nu_{\rm s} = 5 \times 10^6 \gamma^2 B${\,Hz}
(in the blob's frame)
\footnote{
    For a single electron it is $\nu_{\rm s} = 2.8 \times 10^6 \gamma^2 B${\,Hz} (e.g., Tucker 1975).
    For the electron distribution in Eq.(\ref{eq:el_spectr}) the average value of $\gamma$ is
    $\bar \gamma=f(n_1,n_2)\,\gamma{\rm br}$. Typical values of $n_1$, $n_2$ imply $\bar \gamma
    \simeq 1.3\, \gamma_{\rm br}$.
}
and $B \simeq 0.02$\,G in all the eight Mrk\,501 states under consideration, the latter condition translates into
log\,$\gamma \ll 5$. We then have hence
\begin{eqnarray}
{ {\rm log} (\nu_{\rm IC} / \nu_{\rm s}) \over {\rm log} \gamma} \,\,
= \, 2\,, \,\, \,\, \,\, \,\, {\rm log}\gamma \ll 5.0 \,;
\label{eq:Thomson}
\end{eqnarray}
\smallskip

\noindent
{\it (ii)} in the high-frequency limit $\epsilon^\prime \gg m_e c^2$ (extreme K-N regime) it is
$\epsilon_1^\prime = m_e c^2$, so $\epsilon_1 \simeq \gamma m_e c^2$. In the SSC framework we
have, for one electron, $\epsilon_1 = h \nu_{\rm IC}$, so $\nu_{\rm IC} \simeq \gamma m_e c^2
h^{-1}$, and $\nu_{\rm s} = 2.8 \times 10^6 \gamma^2 B${\,Hz}. For the case, of relevance here, of a
broken-PL electron spectrum, it is $\nu_{\rm IC} \simeq \gamma m_e c^2 h^{-1} g(\alpha_1, \alpha_2)$
[with $g(\alpha_1, \alpha_2) = {\rm exp}[(\alpha_1-1)^{-1}+0.5\,(\alpha_2-\alpha_1)^{-1}]$, being
$\alpha_{1,2} = (n_{1,2}-1)/2$] (see T98) and $\nu_{\rm s} = 5 \times 10^6 \gamma^2 B${\,Hz}. For values
$B=0.02\,$G, $n_1=1.7$, $n_2=3.0$, appropriate for Mrk\,501 (see Table\,1), we get
\begin{eqnarray}
{ {\rm log} (\nu_{\rm IC} / \nu_{\rm s}) \over {\rm log} \gamma} \,\, \simeq \,\, { 14.75 \over {\rm log} \gamma} -1\,,
\,\, \,\, \,\, \,\, {\rm log}\gamma \gg 5.0 \,;
\label{eq:KN}
\end{eqnarray}

The predictions in Eqs.({4}) and ({5}) can be graphically tested by plotting (in a log-log plane) the
IC-to-synchrotron peak distance vs. $\gamma$ (see Fig.~\ref{fig:Thomson_KN})%
\footnote{Since in our assumed scheme the dominant synchrotron power is emitted by electrons at
      the break (i.e., with $\gamma_{\rm br}$), and the dominant IC power (in the Thomson
      regime) is accounted for by photons with frequency $\sim \gamma_{\rm br}^2 \nu_{\rm s}$,
      the relevant $\gamma$ in this argument is $\gamma_{\rm br}$. }%
: states 1-5 (quiescent to moderately active) are located on, or close
to, the Thomson line whereas states 6-8 (active) are located close to the extreme K-N curve.

\subsection{Parameter and source evolution}

Examining Table\,1, we may tentatively point out some trends shown by the SSC parameters during the evolution of
an active event. States 6, 7, and 8, snapshots of the active event of April 1997 that culminated in the Giant Flare
(Pian et al. 1998), taken $\Delta t$ days apart from one another, suggest that the quantity $B\,R^2$ is conserved
(as expected in a fully ionized plasma), and -- assuming the rise to the Giant Flare to have started $\Delta t$ days
before state\,6 -- that the blob radius grows with time as $R \propto t^{2/5}$.

If real, the latter suggested regularity corresponds (Wand \& Kusunose 2002) to the self-similar solution (Sedov 1959)
of the subrelativistic expansion of a plasma blob. Pursuing this analogy further, this Sedov expansion leads to the formation
of a strong shock which then sweeps some electrons from the medium and accelerates them to relativistic energies:
so the expansion acts as the injection of relativistic electrons (Wand \& Kusunose 2002).

These elements may suggest the following picture for the early development of a flare: the blob expands while traveling
along the jet, and fills it because its internal radiation pressure swells it until magnetic confinement forces prevail
and push the blob further out where it can {keep expanding}. As long as new particles are accelerated so that the internal
pressure builds up, one can conjecture a positive feedback among expansion, motion along the jet, internal
shock formation, and particle acceleration until maximum emission is reached.

\subsection{Comparison with Mrk\,421}

A cursory comparison with Mrk\,421 (Mankuzhiyil et al. 2011) suggests that Mrk\,501 (see Table\,1) has a larger
(by a factor $\sim$12) emission size $R$; a lower but more stable Doppler factor ($\delta \sim 20 \pm 5$) [whereas
Mrk\,421 has $\delta \sim (25\pm5) - (100\pm20$)]; and a lower $B$. So the two jets appear to be substantially
different from each other.

A deeper insight into similarities and differences in the emission physics of the two jets can be reached by examining
correlations between SSC parameters (e.g., Tavecchio et al. 1998; see Fig.~\ref{fig:SSC_param}). A $B$--$\gamma_{\rm br}$
anticorrelation is predicted if the synchrotron peak, $\nu_{\rm s} \propto B \gamma^2$, remains (roughly) constant from
state to state. For constant $\nu_{\rm s}$ and $\nu_{\rm IC}$, a $B$--$\delta$ relation is predicted to be {\it inverse}
in the Thomson regime and {\it direct} in the {extreme} K-N limit (e.g., Tavecchio et al. 1998). Both $B$--$\gamma_{\rm br}$ and
$B$--$\delta$ inverse correlations do hold for the 9 SEDs of Mrk\,421 examined by Mankuzhiyil et al. (2011). For Mrk\,501
the frequencies $\nu_{\rm s}$ and $\nu_{\rm IC}$ are not constant, hence the $B$--$\gamma_{\rm br}$ (anti)correlation does
not hold (see Fig.~\ref{fig:SSC_param}). Similarly, in the $B$-$\delta$ plot there seems to be no correlation for Mrk\,501,
possibly suggesting a mix of (predominantly) Thomson and (predominantly) {extreme} K-N states for this source. More directly, from
Fig.~\ref{fig:Thomson_KN} we see that all Mrk\,421 points are on, or close to, the Thomson line, where the synchrotron and
Compton peak frequencies are related by $\nu_{\rm IC}/\nu_{\rm s} \propto \gamma^2$, whereas the Mrk\,501 points are found
close to {\it both} the Thomson and the {extreme} K-N curves.

Using the best-fit SSC parameters, for each SED of both galaxies we compute the electron energy density $U_e$
and the magnetic energy density $U_B$. The results (see Fig.~\ref{fig:SSC_param}) show that the the situation
is far from particles-field equilibrium, with the electrons dominating over the field by orders of magnitude.
\footnote{
    In both sources $U_e$ is computed assuming $\gamma_{\min}=1$.
    The impact of $\gamma_{\min}$ on $U_e$ is not dramatic as long
    as $n_1 \leq 2$, which is the case here. So our conclusion on
    particle dominance, although based on one particular value of
    $\gamma_{\min}$, retains general validity.
}
(If we {would also consider a contribution from} accelerated protons, which appreciably contribute not to the radiative yield but to the
mechanical power of the jet, our conclusion is even stronger.) These results suggest that in all the states
considered in this paper, quiescent and active, the jets of both Markarians seems to be primarily kinematic
-- as opposed to electromagnetic. This result was already pointed out, for quiescent states alone,
by e.g. Kino et al. (2002). Both the magnetic and the electron energy densities are distinctly higher
in Mrk\,421 than in Mrk\,501; however Mrk\,421 suggests a $U_B \propto U_e^2$ behavior, whereas no correlation
appears for Mrk\,501 -- especially without the two lowest-$U_B$, highest-$U_e$ points that correspond to the
flaring states 5 and 8 (see Fig.~\ref{fig:SSC_param}).

\section{Summary}

Through $\chi^2$-minimization we have modeled eight simultaneous SED datasets of the HBL source
Mrk\,501 with a one-zone SSC model. Our results suggest that Mrk\,501's variability is complex,
with the IC component being produced by photon-electron scatterings that at times are in prevalent
Thomson regime and at times in prevalent extreme K-N regime. The variations of the SED have enabled us to
track the corresponding variations of the underlying emitting electron spectrum: from distinctly
broken-PL and less intense and softer in the non-flaring states, to virtually single-PL and more
intense and harder at the peak of flares. The former are interpreted as the steady-state spectra of
aging electrons, the latter as the fresh electron injection spectrum. Comparing data for Mrk\,501
and Mrk\,421, the two jets mainly differ by the physical characteristics of their emission regions,
but their energy budgets appear to be both particle dominated.

\section*{Acknowledgements}

{We acknowledge sensible comments from an
anonymous referee that have led to a substantial
improvement of this paper.}

\begin{table*}
\begin{center}
{\footnotesize%
\begin{tabular}{l|cccccccc}
\hline
\hline
\hline
\noalign{\smallskip}
        & State 1    & State 2       & State 3     & State 4        & State 5        & State 6         & State 7   & State 8  \\
\noalign{\smallskip}
\hline
\hline
Date    & March           & Mar-Aug         & Jul          & 15-26 Jun      & 27-28 Jun      & 7 Apr           & 11 Apr        & 16 Apr  \\
        & 2009            & 2009            & 2006         & 1998           & 1998           & 1997            & 1997          & 1997    \\
Instr.  &                 & Swift           & KVA          &                &                &                 &               &         \\
        & {\it Suzaku}    & {\it R}XTE      & {\it Suzaku} & {\it R}XTE     & {\it R}XTE     & {\it Beppo}SAX  &{\it Beppo}SAX & {\it Beppo}SAX \\
        & {\it Fermi}/LAT & {\it Fermi}/LAT &              &                &                &                 &               &         \\
        & MAGIC           & MAGIC           & MAGIC        & HEGRA          & HEGRA          & CAT             & CAT           & CAT     \\
        & VERITAS         & VERITAS         &              &                &                &                 &               &         \\
Ref.    & [1]             & [2]             & [3]          & [4]            & [4]            & [5]             & [5]           & [5]     \\
\noalign{\smallskip}
\hline
\hline
\noalign{\smallskip}
Param. &  &  &  &  &  &  &  &  \\
\noalign{\smallskip}
\hline
\noalign{\smallskip}

$K_e$                  &$68.9^{0.2}_{0.2}$      &$234^{2}_{2}$           &$254^{2}_{2}$           &$153^{1}_{1}$           &$456^{4}_{4}$           &$94^{1}_{1}$          &$165^{1}_{1}$    & $465^{4}_{4}$  \\
$\gamma_{\rm{min}}$  &$1$                     &$1$                     &$1$                     &$1$                     &$1$                     &$1$                    &$1              $  & $1       $  \\
$\gamma_{\rm{br}}$   &$8.02^{0.01}_{0.01}$    &$4.88^{0.03}_{0.03}$    &$5.60^{0.02}_{0.02}$    &$2.67^{0.02}_{0.02}$    &$1.03^{0.01}_{0.01}$    &$26.72^{0.02}_{0.02}$  &$30.64^{34.9}_{30.0}$    & $55.5^{2.1}_{1.9}$  \\
$\gamma_{\rm{max}}$  &$1.86^{0.04}_{0.03}$    &$2.12^{0.06}_{0.05}$    &$1.61^{0.22}_{0.15}$    &$2.67^{0.08}_{0.07}$    &$3.62^{0.06}_{0.06}$    &$12.73^{0.05}_{0.05}$  &$13.53^{13.3}_{3.3}$    & $8.85^{0.36}_{0.30}$  \\
$n_{1}$              &$1.73^{0.00}_{0.00}$    &$1.79^{0.00}_{0.00}$    &$1.78^{0.00}_{0.00}$    &$1.73^{0.00}_{0.00}$    &$1.64^{0.00}_{0.00}$    &$1.65^{0.00}_{0.00}$   &$1.70^{0.00}_{0.00}$      & $1.73^{0.00}_{0.00}$  \\
$n_{2}$              &$3.38^{0.00}_{0.00}$    &$3.16^{0.00}_{0.00}$    &$3.61^{0.00}_{0.00}$    &$3.05^{0.00}_{0.00}$    &$2.25^{0.02}_{0.02}$    &$3.00^{0.01}_{0.01}$   &$2.80^{0.00}_{0.00}$     & $2.11^{0.01}_{0.01}$  \\
$B$                  &$2.098^{0.002}_{0.002}$ &$1.530^{0.006}_{0.006}$ &$5.58^{0.02}_{0.01}$ &  $1.513^{0.005}_{0.005}$  &$1.126^{0.006}_{0.006}$ &$3.60^{0.02}_{0.02}$&$1.799^{0.009}_{0.008}$    & $1.043^{0.006}_{0.005}$  \\
$R$                  &$2.332^{0.002}_{0.002}$ &$1.909^{0.006}_{0.006}$ &$1.620^{0.004}_{0.004}$ &$1.985^{0.005}_{0.005}$ &$1.742^{0.006}_{0.006}$ &$0.925^{0.003}_{0.003}$&$1.202^{0.004}_{0.003}$   & $1.627^{0.004}_{0.004}$  \\
$\delta$             &$19.13^{0.01}_{0.01}$   &$24.40^{0.05}_{0.05}$   &$15.12^{0.02}_{0.02}$   &$25.24^{0.05}_{0.05}$   &$12.85^{0.03}_{0.03}$   &$16.43^{0.04}_{0.04}$  &$17.53^{0.04}_{0.04}$  & $13.89^{0.03}_{0.03}$ \\

\hline
log$\,L$               & 44.57      & 44.58      & 44.59     & 44.54     & 44.91      & 44.95     & 45.07     & 45.53\\
log$\,\nu_{\rm s}$     & 16.74      & 16.86      & 16.34     & 17.16     & 18.39      & 19.30     & 19.51     & 19.27\\
log$\,\nu_{\rm IC}$    & 25.36      & 25.46      & 24.88     & 25.34     & 26.35      & 26.17     & 26.53     & 26.93\\

\noalign{\smallskip}

\hline
\hline
\hline
\end{tabular}}
\end{center}
\caption{
Datasets and best-fit single-zone SSC model parameters for the eight datasets of Mrk\,501.
States are named as in Fig.1. References are: [1] Acciari et al. (2010), [2] Abdo et al. (2011), [3] Anderhub
et al. (2009), [4] Sambruna et al. (2000), [5] Djannati-Ata\"{\i} et al. (1999).
$B$ is given in units of $10^{-2}$G, $R$ in units of $10^{16}$cm, $K$ in units of cm$^{-3}$, $\gamma_{\rm br}$ in
units of $10^4$, $\gamma_{\rm max}$ in units of $10^6$. The isotropic luminosities (in erg\,s$^{-1}$,
given in log), and the synchrotron and IC peak frequencies (in Hz, given in log), are derived from the SEDs'
best-fit SSC models. }
\label{tab:results}
\end{table*}

\end{document}